\documentstyle[eqsecnum,aps,epsf]{revtex}
\begin{document}

\title{A study of the phase transition in the transverse Ising model 
using the extended coupled-cluster method}
\author{J. Rosenfeld}\address{IRST-SSI, Loc. Pante' di Povo,  I-38100 Povo (Trento), Italy} 
\author{N.E. Ligterink
\footnote{e-mail: ligterin@ect.it}}
\address{ECT*, Villa Tambosi, Strada delle Tabarelle 286,
I-38050 Villazzano (Trento), Italy}
\date{\today}
\maketitle

\begin{abstract}
The phase transition present in the linear-chain and square-lattice 
cases of the transverse Ising model is examined.
The extended coupled cluster method (ECCM) can describe
both sides of the phase transition with a unified approach. 
The correlation length and the excitation energy are determined.
We demonstrate the ability of the ECCM to use both the weak- and the 
strong-coupling starting state in a unified approach for the study 
of critical behavior.
\end{abstract}


\section{Introduction}
The study of the magnetic properties of materials from microscopic theory 
has always proved difficult. Even at the level of such simplified 
magnetic models as the Heisenberg model, 
the Ising model, and the
Hubbard model, intensive theoretical and numerical work is required to 
model the interesting counter-intuitive behavior they exhibit.

An obstacle to studying quantum magnets, 
encountered by various theoretical approaches, is due to the uniqueness 
of the state at the phase transition point. 
Most techniques can describe a particular phase 
of a quantum magnet but may be unable to describe the
other phase, or phases, of the system and, hence, the phase transition itself.
Clearly, a consequence for such techniques is that conclusive evidence of 
symmetry breaking cannot be found. 

All techniques dealing with many-body systems experience the difficulty of 
being able to include sufficient correlations,
without making prior assumptions about the leading correlations, 
in order 
to provide a good description of the important behavior of the system.
Unsurprisingly, the subtleties of a 
quantum magnet at the phase transition, which require the inclusion of 
a great many correlations to model adequately,
are notoriously difficult to obtain. 
Many theoretical approaches \cite{US} yield a mean-field like description of the 
system, which does not reveal the distinctly different behavior present in 
differing dimensions, contrary to the true physical behavior.  \cite{MWC}

In this paper, we present a study of the transverse Ising model via 
an improved approach \cite{thesis,Josi2} recently developed for 
spin systems. This approach involves the application of 
the quantum many-body technique known as the extended coupled-cluster
 method (ECCM). {\it A priori}, we know that the ECCM is potentially 
a good technique for studying the phase transition as it possesses certain 
features that allow it to study spontaneous symmetry breaking. In particular, 
the ECCM can yield a single solution that correctly describes the 
physical behavior on either side of the phase transition.
The more widely applied normal coupled-cluster method (NCCM), 
which is a restricted version of the ECCM, at a given level of 
approximation, does not possess this feature of the ECCM. A comparison 
of the results from these techniques will demonstrate 
the superiority of the ECCM 
with regard to the study of the phase transition. Most importantly, we
 aim to show that the improved approach outlined 
in \cite{Josi2} yields trustworthy and accurate numerical data that can 
provide real insight into the physical behavior of the system.

The transverse Ising model has two distinct phases; 
the ferromagnetic phase, and the paramagnetic phase,
which are tuned by a single coupling constant. On the linear chain, this model
has some interesting properties, such as the duality, which allows one to
relate the ground-state energy at a particular value of the coupling constant
to the ground-state energy at the reciprocal value of the coupling constant.
This property and the fact the model is exactly solvable on the linear 
chain makes it a good test-ground for the ECCM.  With the confidence gained 
from the linear-chain model, we will then study the square-lattice case.

\section{The Transverse Ising Model}

The transverse Ising Hamiltonian has two competing terms on a lattice
composed of spin-half spins; the ferromagnetic term,
which forces the system to align in the $z$-direction, and the paramagnetic
term, which forces the system to align along the positive $x$-axis:
\begin{equation}
H =  - \sum^{d N}_{\langle {\bf i}, {\bf j} \rangle}  \sigma^z_{\bf i}
 \sigma^z_{\bf j} 
- \lambda \sum^N_{\bf i} \sigma^x_{\bf i} \ \ ,
\end{equation}
where $\sigma^i$ are the Pauli matrices, 
$\lambda $ is the external magnetic field, $N$ is the number 
of lattice sites,
$d$ the dimensionality of the lattice, and  the sum over 
$\langle {\bf i}, {\bf j} \rangle$ runs over all nearest-neighbor
 pairs 
of lattice sites ${\bf i}$ and ${\bf j}$.
In the weak-coupling  limit, $\lambda = 0$, the spins will align themselves
along the $z$-axis, in either the positive or the negative direction. 
In the strong-coupling limit, $\lambda \to \infty$, the magnetic field forces 
the spins to align along the positive $x$-axis. Between these two
phases a phase transition occurs for some critical coupling $\lambda_c$.

Starting from the strong-coupling limit, the $Z(2)$ symmetry is preserved,
which results in the $\langle \sigma^z \rangle =M^z$ being
zero. Beyond the critical point this symmetry
is broken, and two degenerate ground states exist, with $\pm M^z$.
Any {\it ab-initio} method would have difficulties in breaking such a symmetry. 
We will show that within a specific approximation, we can start with
an uncorrelated model state in both limits, which can either incorporate
the $Z(2)$ symmetry or break it, in the latter case two  exactly degenerate 
states exist.

\subsection{Exact results on the linear chain}
On the linear chain the model is exactly solvable. \cite{LSM61,CDS96}
After a Jordan-Wigner transformation
the Hamiltonian can be diagonalized using a Bogoliubov transformation. 
The ground-state energy is given by:
\begin{eqnarray}
\frac{E_g(\lambda)}{N}  & = & 
-\frac{1}{\pi} \int_0^{\pi} d q  {\sqrt{1+2\lambda \cos q + \lambda^2}}  \cr
& = &  - \frac{2}{\pi}(1+\lambda) {\cal E}\left( \frac{4 \lambda}{(1+\lambda)^2}
 \right)\ \ ,
\end{eqnarray}
where ${\cal E}$ is a complete elliptic integral of the second kind.
This energy satisfies the interesting duality, as could be concluded 
from the duality of the original Hamiltonian:
\begin{equation}
 E_g(\lambda) = \lambda E_g(\lambda^{-1}) \ .
\end{equation}
The magnetization $M^{\alpha} = \langle \sigma^{\alpha} \rangle$ 
is given by:
\begin{eqnarray}
M^x & =  & 
\frac{1}{\pi} \int_0^{\pi} \frac{(\lambda + \cos q) d q}
{\sqrt{1+2\lambda \cos q + \lambda^2}} \ , \\
 M^z & = & \theta(1-\lambda)\left(1-\lambda^2 \right)^{\frac{1}{8}} \ ,
\end{eqnarray}
where $\theta (x)$ is the step function; $\theta(x<0) =0 $ and $\theta (x>0) = 1$.
The transverse magnetization $M^x$ can be related to the 
derivative of the ground-state energy, however, the longitudinal 
magnetization $M^z$ is derived from  the limit
of the spin-spin correlation function with a lattice size taken 
to infinity,
therefore the vanishing long-range order for $\lambda >1$ 
is a subtle result.\cite{LSM61}

\subsection{The mean field}
\label{mfc}
A parameterization of the wave function is given by:
\begin{equation}
| \Psi \rangle = \left.\left| 
\left( \matrix{ -\sin \theta/2 \cr \ \ \ \cos \theta/2 } \right),
\left( \matrix{ -\sin \theta/2 \cr \ \ \ \cos \theta/2 } \right),
\left( \matrix{ -\sin \theta/2 \cr \ \ \ \cos \theta/2 } \right),
\left( \matrix{ -\sin \theta/2 \cr \ \ \ \cos \theta/2 } \right),
 \cdots \right.\right\rangle \ \ .
\end{equation}
From the variational principle the mean-field solution 
yields:
\begin{eqnarray}
\theta & = &  -\arcsin\bar\lambda\ \  ;  \ \ 
\bar\lambda\leq 1\ \ ,\\
 & = &  - \frac{\pi}{2} \ \  ; \ \ \bar\lambda\geq 1\ \ ,\\
\frac{E_g}{N} & = &  -d\left(1+\bar\lambda^2\right)\ \ ;\ \ 
\bar\lambda\leq 1\ \ ,\\
              & = &  - 2d \bar\lambda \ \  ; \ \ \bar\lambda\geq 1\ \ ,\\
M^z & = &  \theta\left(1-\bar\lambda \right)\sqrt{1-\bar\lambda^2}\ \   ,
\end{eqnarray}
where $\lambda = 2d \bar \lambda$. 
Note that $\theta \to -\pi - \theta$ is also a solution with the same energy 
but opposite magnetization. This is the $Z(2)$ symmetry partner of the 
ferromagnetic ground state.

\section{THE EXTENDED-COUPLED CLUSTER METHOD}

The Coupled-Cluster Method (CCM) which exists in two distinct 
versions, the NCCM and the ECCM, is a powerful technique for 
quantum many-body calculations. Only the 
essential features of the CCM will be reviewed here, 
further details can be found in the literature. \cite{Arp,Bishop,Josi2}
There are four key features of the CCM. 
Firstly, the model state $|\Phi\rangle$, a uncorrelated state,
which is usually the true ground state in a certain limit, serves as the starting
state. Secondly, the correlations are incorporated in the the ket
state in the exponentiated form:
\begin{equation}
|\Psi \rangle=e^S|\Phi\rangle\ \ ,
\label{ket}
\end{equation}
where the correlation
operator $S$ is composed solely of a
complete set of mutually commuting
multiconfigurational creation operators that annihilate the 
bra model state $\langle\Phi|$. 
The exponential form ensures the correct summation of
the independent correlations in the calculation, and spreads
correlations over the whole system although the
correlations in $S$ might be only local.

The third key feature of the CCM appears in the parameterization of
 the bra state $\langle \tilde \Psi |$,
\begin{equation}
\langle \tilde \Psi |  =   \langle \Phi | (1 + \tilde S) e^{- S} =
 \langle \Phi |  e^{S^{''}} e^{- S} \ \ ,
\label{parNE}
\end{equation}
where the first parameterization gives rise to
 the NCCM, and the second parameterization gives rise to the ECCM. \cite{Arp}
The importance of both of the above bra-state parameterizations are
 that they allow the similarity transform to be incorporated into the 
CCM calculation of any observable.
 Primarily, the similarity transform only yields
 terms for which the correlations present in $S$ are linked 
to the observable, and, hence, subsequent expectation values are 
size-extensive quantities. Importantly, the presence of the similarity 
transform means that no further truncations are necessary than the truncations
of the number of configuration of creation operators in $S$.
In this way the similarity transform can be seen as the defining 
feature of the CCM, since its absence would render an approximate 
calculation impractical. For example, if one would attempt a variational
calculation with the ket state of Eq.~(\ref{ket}) and its Hermitian conjugate,
several further approximations are required to calculate the functional.
Therefore, the actual ket state obtained from the same parameterized variational
wave functional depends on the associated bra state and the corresponding 
variational principle.

This brings us to the fourth key feature of the CCM, which is related to the 
parameterization of the 
bra state in Eq. (\ref{parNE}) and concerns the terms arising from 
$ e^{S^{''}}$ and 
${\tilde S}$. These terms are present in the ECCM and NCCM respectively and
are solely composed of destruction operators; the Hermitian conjugates
of the creation operators in $S$. Therefore, the functional 
$\langle \tilde \Psi | H | \Psi \rangle$ is a polynomial in the
bra and ket coefficients, associated with the strength of the different
creation and annihilation operators in $S$ and 
$S^{''}$ respectively. \cite{Arp}

Both of the NCCM and ECCM
parameterizations result in the maintenance of some useful 
formal properties, which remain valid in any approximation scheme, 
e.g., Feynman-Hellman theorem and the linked-cluster theorem for the ket
state coefficients. \cite{Bishop}
However, the ECCM, unlike the NCCM, parameterization 
retains, in any approximation scheme, the full cluster separability:
\begin{equation}
\lim_{{\bf | r - r^{'} |}\rightarrow\infty}
\langle {A_{{\bf r}}  B_{\bf r^{'}}}\rangle =
\langle{{A}_{\bf r}}\rangle\langle{{B}_{\bf r^{'}}}\rangle\ \ ,
\label{global}
\end{equation}
where $A_{\bf r}$ and $B_{\bf r^{'}}$ are  localized
operators acting at positions ${\bf r}$ and ${\bf r^{'}}$, 
respectively.  Therefore, it allows us to separate the correlation
spin-spin function $\langle \sigma^x_{\bf r} \sigma^x_{\bf r'}\rangle$
into a size-extensive part, the long-range order; 
$\langle \sigma^x\rangle^2=(M^x)^2$, and
the true correlation function that vanishes at infinite distance.

\subsection{Transformation of the Hamiltonian in terms of a canted model state}

For a unified description of the different model states $|\Phi \rangle$,
we perform a unitary transformation such that the model state, which is
a mean field state where the spins point in a particular direction, is a
state composed of down-pointing spins.
The details of this transformation have been 
given elsewhere. \cite{thesis,Josi2}
Hence, the creation operators in $S$ will be products 
of $\sigma^+$'s at different
sites.
Using the rotation matrix $U$, for this purpose,
 defined by,
\begin{equation}
U\equiv \exp\left(i\theta\frac{\sigma^y}{2}\right) 
 = \cos\left(\frac{\theta}{2}
\right) \openone + i\sin\left(\frac{\theta}{2}\right)
 \sigma^y\ \ ,
\label{ub}
\end{equation}
one can obtain the transverse Ising Hamiltonian defined with respect 
to an arbitrary canted model state, in terms of rotated operators corresponding
to the down-spin model state. 
An illustration of the particular 
canted states of the ferromagnetic and paramagnetic states, described by the angle 
$\theta$, are respectively given by:
\begin{equation}
\theta =0\ \ ; \ \ |\downarrow\downarrow\downarrow\downarrow\cdots
\rangle_{\lambda = 0}\ \ ,\ \  \theta=-\pi/2\ \ ;\ \ 
|\rightarrow\rightarrow\rightarrow\rightarrow\cdots
\rangle_{\lambda\to \infty}\ \ .
\end{equation}
The rotated transverse Ising Hamiltonian is of the form:
\begin{eqnarray}
UHU^{\dagger}& = &-\sum_{\langle {\bf i,j}\rangle}^{dN}\left[\sin^2\theta\sigma_{\bf i}^x
\sigma_{\bf j}^x +\cos^2\theta\sigma_{\bf i}^z\sigma_{\bf j}^z+
\sin2\theta \sigma_{\bf i}^{x}\sigma_{\bf j}^z \right] \cr
&&-\lambda \sum_{\bf i}^N\left[-\sin\theta\sigma_{\bf i}^z+
\cos\theta\sigma_{\bf i}^x\right]\ \ .
\end{eqnarray}
In most cases we use only two rotations: the ferromagnetic
Hamiltonian, with $\theta=0$, and the paramagnetic Hamiltonian, with 
$\theta= -\frac{\pi}{2}$.

\subsection{The SUB1 scheme and symmetry breaking}
The ECCM SUB1 approximation scheme only retains 
the one-body correlations present in $S$ and 
$S^{''}$ in 
Eqs. (\ref{ket}) and (\ref{parNE}) defined with respect to the 
rotated model state, given by:
\begin{equation}
S = k\sum_{\bf i}\sigma^+_{\bf i}\ \ ;\ \ 
S^{''}=k^{''}\sum_{\bf i}\sigma^-_{\bf i}\ \ .
\end{equation}
Performing the CCM SUB1 scheme is equivalent to 
a mean-field calculation with an important
 proviso. Crucially, the ECCM SUB1 scheme can yield the full 
mean-field ground-state solution, however,
 the NCCM SUB1 scheme 
can not yield the ground-state solution for the region beyond the phase transition. This feature of the NCCM 
has dire implications for the study of 
symmetry breaking. It is a little known fact that, in practice,
 the CCM yields a solution 
that preserves the shared symmetry of the model state and the Hamiltonian \cite{Yang}. Fortunately, 
via the use of the ECCM SUB1 scheme, it 
is possible to obtain a symmetry-broken 
solution in a correlated ECCM calculation.

The relationship between the 
mean-field ground state and the ECCM 
one-body coefficients 
defined with respect to the canted model 
state, where the spin are rotated downwards,
 is given by: \cite{thesis}
\begin{equation}
k=-\tan\left(\frac{\theta}{2}\right)\ \ ;\ \ 
k^{''}=-\frac{1}{2}\sin\theta\ \ . 
\label{kval}
\end{equation}
In a correlated ECCM calculation, 
these relations are only valid in the 
classical regimes, $\lambda=0$ and $\lambda=\infty$, of the system.
By using the values of $k$ and $k^{''}$, from 
Eq. (\ref{kval}), starting
from an initial point in the classical region 
of the phase where the chosen model state is 
not the classical wave function, one can determine 
the symmetry-broken solution.

The above ansatz is put into practice
to study the possibility of broken 
$Z(2)$ symmetry in the transverse Ising model. 
Employing the paramagnetic model state, with the values 
$k=1$ and $k^{''}=1/2$,  which has the advantage of not
artificially breaking the $Z(2)$ 
symmetry, 
corresponds to the ferromagnetic state. 
At the 
ferromagnetic point $\lambda=0$ it  yields the 
symmetry-broken solution. For $k=0$ and 
$k^{''}=0$ the symmetry preserved solution 
can be determined, for which $M^z=0$, from the limit $\lambda \to \infty$.

\subsection{The diagrammatic representation of the SUB2-$n$ scheme functional}
In order to perform the numerical ECCM calculation 
efficiently, a diagrammatic \cite{US,Josi2} representation is 
implemented. The ECCM SUB2-$n$ approximation scheme is implemented here and is described in greater 
depth in  References \cite{Arp,Josi2}. Essentially, the 
SUB2-$n$ scheme retains only one and 
two-body correlations, where 
the truncated correlation operators are of the 
form: 
\begin{equation}
S = k\sum_{\bf i}\sigma^+_{\bf i} +
\sum_{\bf i,\bf r}^{\chi\leq n} b_{\chi({\bf r})}\sigma^+_{\bf i} 
\sigma^+_{\bf i+ r} 
\ \ ,\ \ S^{''} = k^{''}\sum_{\bf i}\sigma^-_{\bf i} +
\sum_{\bf i,\bf r}^{\chi\leq n} b^{''}_{\chi{(\bf r)}}\sigma^-_{\bf i} 
\sigma^-_{\bf i+ r} 
\ \ ,
\end{equation} 
where $\chi({\bf r})$ yields a label for each vector distinct under 
lattice symmetries.

\begin{figure}[h]
\epsfxsize=14cm 
   \centerline{\epsffile{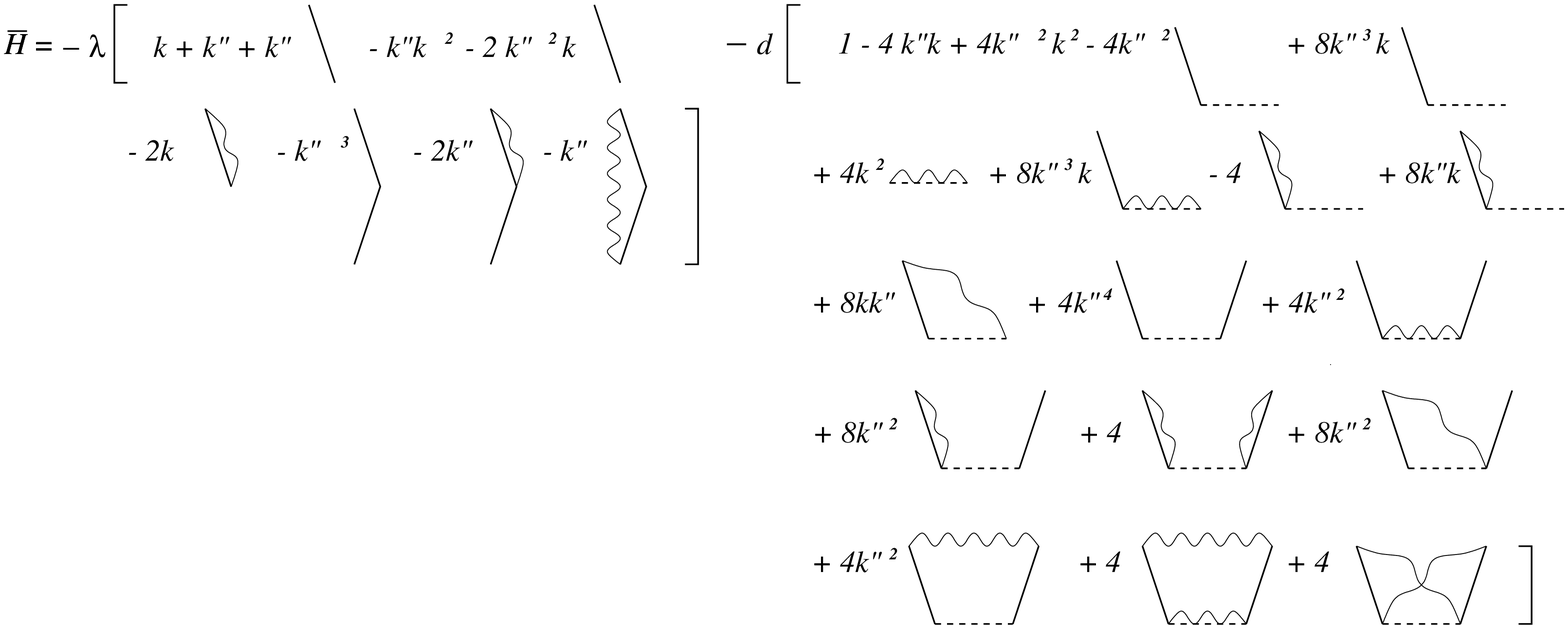}} 
   \caption{The diagrams arising from the transverse Ising model functional 
with the ECCM SUB2-$n$ scheme, which yields the ground-state energy, where
 $d$ is the dimensionality of the hyper-cubic lattice and $\lambda$ 
is the external magnetic field.}
\label{trans}
\end{figure}

The representation of the combinations of 
operators which give rise to various physical quantities is 
outlined in Reference \cite{Josi2}. The transverse Ising functional,  
defined with respect to the ferromagnetic model state
arising from the ECCM SUB2-$n$ scheme is diagrammatically 
expressed in Fig. \ref{trans}.

\subsection{The correlation length and the excitation energies}
Although the order of the approximation schemes is
not sufficient to study the long-distance behavior
of the correlation function, we can examine the correlation
length:
\begin{equation}
 \xi = \frac{1}{N}\langle \sum_{\bf i, j} \sigma^x_{\bf i} \sigma^x_{\bf j} \rangle -
\frac{1}{N^2}\langle \sum_{\bf i} \sigma^x_{\bf i} \rangle^2 \ \ .
\end{equation}
A similar correlation length for the $z$-components of the spins
is computationally too involved to be of any practical use.
This expression for the $\xi$ is easily motivated in the linear-chain
case, where 
\begin{equation}
\langle \sigma^x_{\bf i} \sigma^x_{\bf j} \rangle 
\propto \frac{1}{2} e^{-|{\bf i-j}|/\zeta} + (M^x)^2
\end{equation}
where $M^x$ is the long-range order,
such that $\xi = (e^{\frac{1}{\zeta}}-1)^{-1}\sim \zeta$.

The excitation energies are calculated using the random phase
approximation (RPA), which means that we assume that the excitations
are small, harmonic fluctuations around the ground-state solution.
This leads to a linear eigenvalue problem, \cite{Arp} which is solved 
numerically. We can study the RPA spectrum, since the second-order
derivatives are available for the functional, which is stored 
quasi-analytical for the high-order numerical calculations.
At the phase transition the excitation energy goes to zero for
a second-order phase transition, however,
we know from NCCM \cite{thesis} that the vanishing excitation energy
is also the generic behavior at a termination point.
Similar behavior could occur for ECCM.

\section{Numerical Results}

Usually, there is not much to be gained  from studying the 
ground-state energy in
great detail with the purpose of examining the performance of a method. Its convergence 
is much faster than that of any other quantity, and generally it is not
an indication of the accuracy  of the ground-state wave function.
Moreover, for the transverse Ising model there is no value $\lambda$,
where the numerical value of the energy is of specific interest. 
The position of the phase transition, $\lambda_c$, the only candidate
for a point of interest, is not exactly known. 
The interested reader can contact us for precise numerical values for the 
ground-state energy.
Instead, we focus on the excitation energy and the magnetization.

It is important to note that there are four specific cases studied here. Two different model
states: the symmetry-broken, down-pointing ferromagnetic model state and
the symmetry-preserved, sideways-pointing paramagnetic model state.
These are two distinct approaches to the problem, which lead to two 
different numerical calculations.
However, both can be studied, starting from both of the classical limits, 
$\lambda=0$ and $\lambda = \infty$. The one-body terms, from the SUB1 
calculation, allow us to rotate each state from the model state to the
starting state, which leads to some unexpected differences, which
are summarized in Table
~\ref{re_tab}. 

The ferromagnetic model state, which breaks the $Z(2)$ symmetry, generally
does much worse that the paramagnetic model state. 
We consider the two classical states that can arise with the paramagnetic 
model state. Firstly, the starting state at $\lambda = 0$ breaks the 
$Z(2)$ symmetry since it has $M^z=1$. However, following this
solution for the ground state for increasing
$\lambda$, it turns back and connects up with
the symmetry partner; $M^z \to - M^z$, and ends at the starting
value $\lambda=0$, but now at with $M^z=-1$. At low orders this is a smooth
transition, where $M^z$ goes to zero, however, at high orders it jumps
at a finite value of $M^z$ to $-M^z$. 
At this point the excitation energy vanishes, for any order of approximation.
Therefore both symmetry partners of the ground state are exactly degenerate 
and part of one continuous solution.

\begin{figure}[h]
   \epsfxsize=9cm 
   \centerline{\epsffile{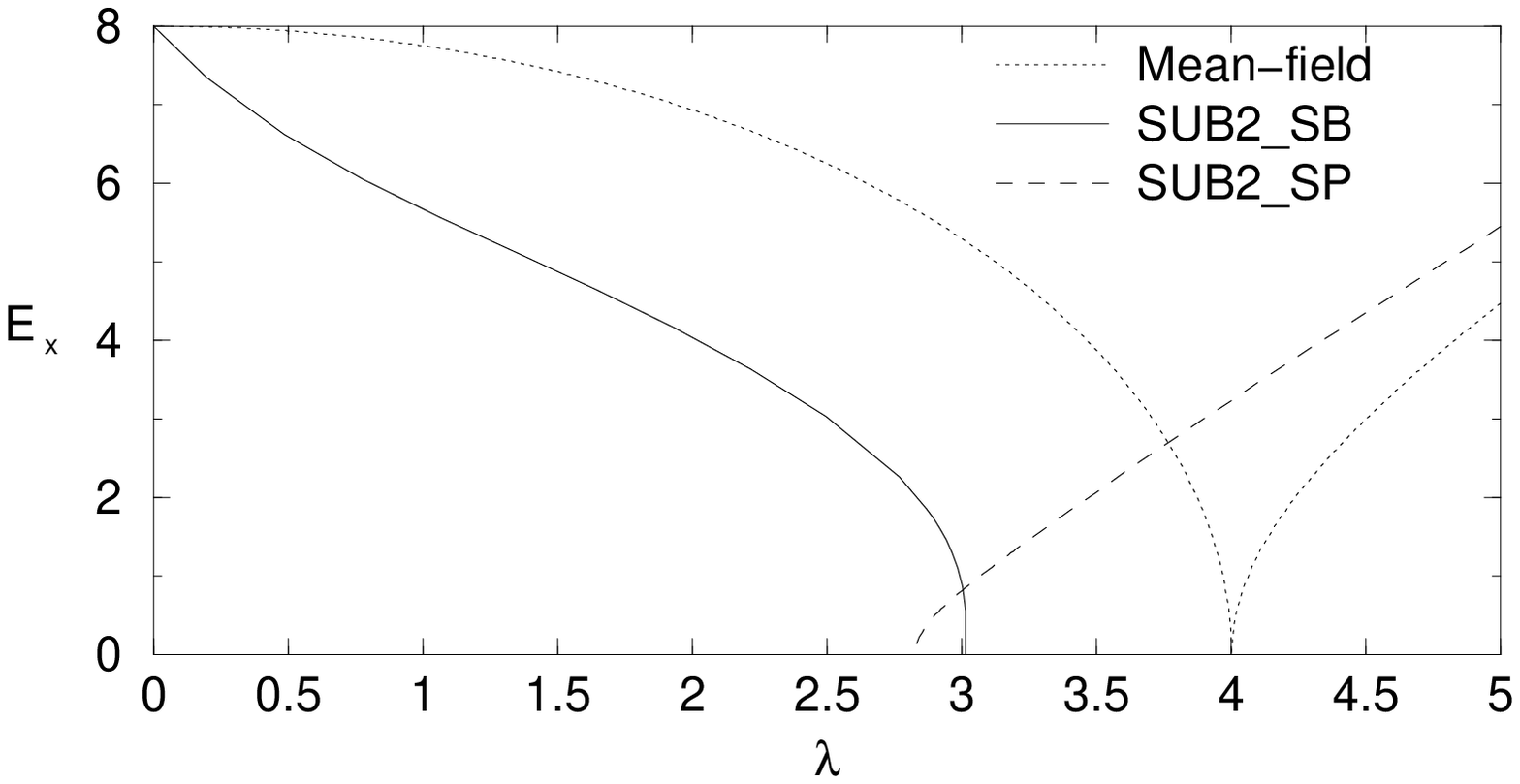 }} 
   \caption{The excitation energy, $E_x$, 
as a function of the transverse magnetic field, $\lambda$, on the 
square-lattice case from the ECCM 
SUB2-$n$ scheme with the paramagnetic model state and
 a mean-field calculation.
}
\label{Exc_2D}
\end{figure}

\begin{figure}[h]
   \epsfxsize=9cm 
   \centerline{\epsffile{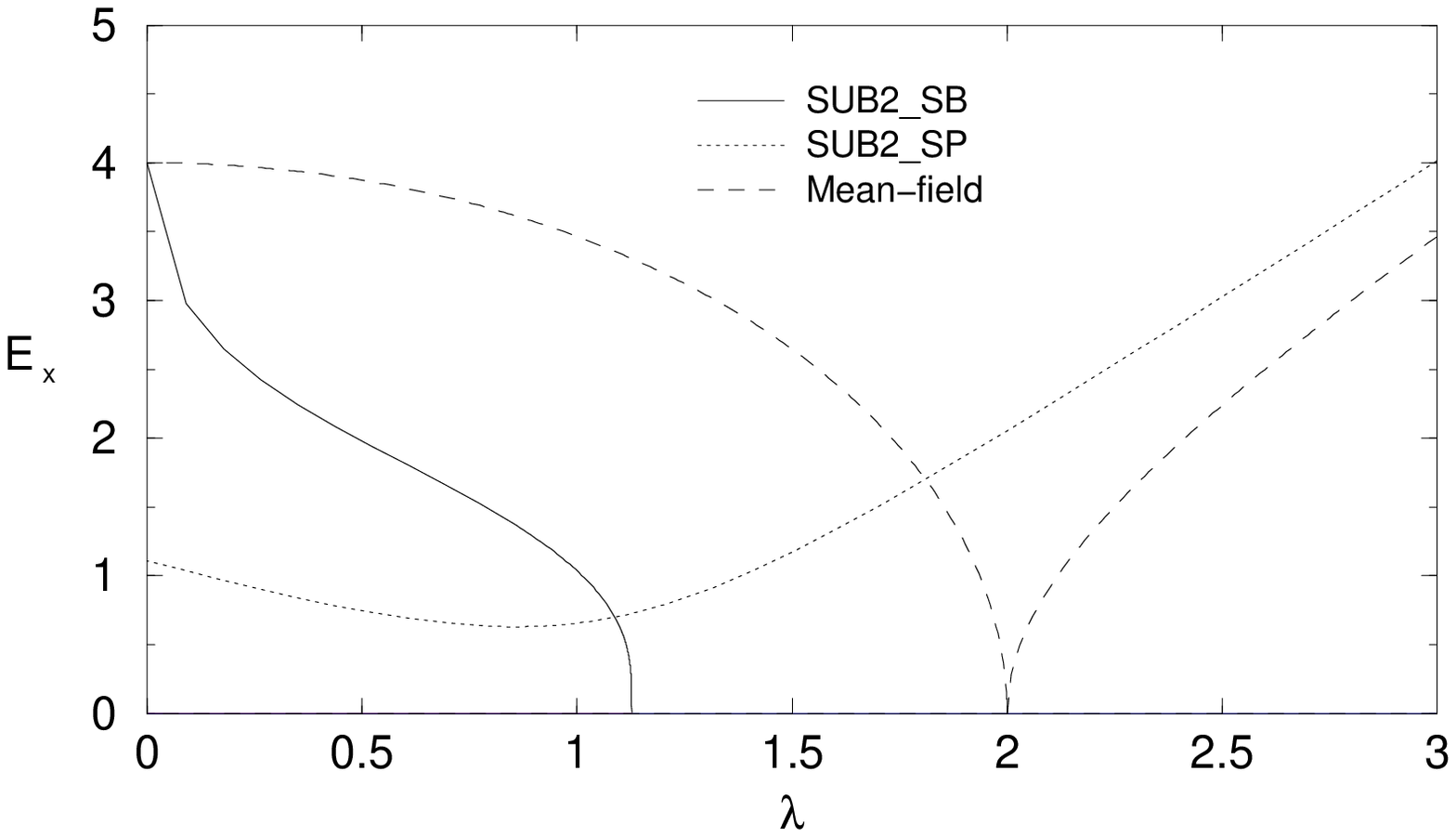 }} 
   \caption{The excitation energy, $E_x$, 
as a function of the transverse magnetic field, $\lambda$, on the 
linear-chain case from the ECCM 
SUB2-$n$ scheme with the paramagnetic model state and
 a mean-field calculation.
}
\label{Exc_1D}
\end{figure}

Secondly, starting with the paramagnetic model state at large values of $\lambda$,
we find that the $M^z$ remains zero, so the symmetry is never broken.
This solution crosses the symmetry-broken solution as  the latter turns back at
$M^z=0$ and $\lambda_c$.
However, looking at the excitation energy of the symmetry preserving
solution, it tends to zero on the
square lattice (see Fig.~\ref{Exc_2D}), but remains finite on the linear chain
(see Fig.~\ref{Exc_1D}), signaling
different behavior. Moreover, the ground-state solution continues beyond the point
where the excitation energy goes to zero, therefore this behavior is
not associated with the possible generic behavior at the termination point.
The value of $\lambda_c$, where the excitation energy vanishes, is 
lower than the termination point of the solution with the paramagnetic model 
state starting from $\lambda=0$.

\begin{figure}[h]
   \epsfxsize=9cm 
   \centerline{\epsffile{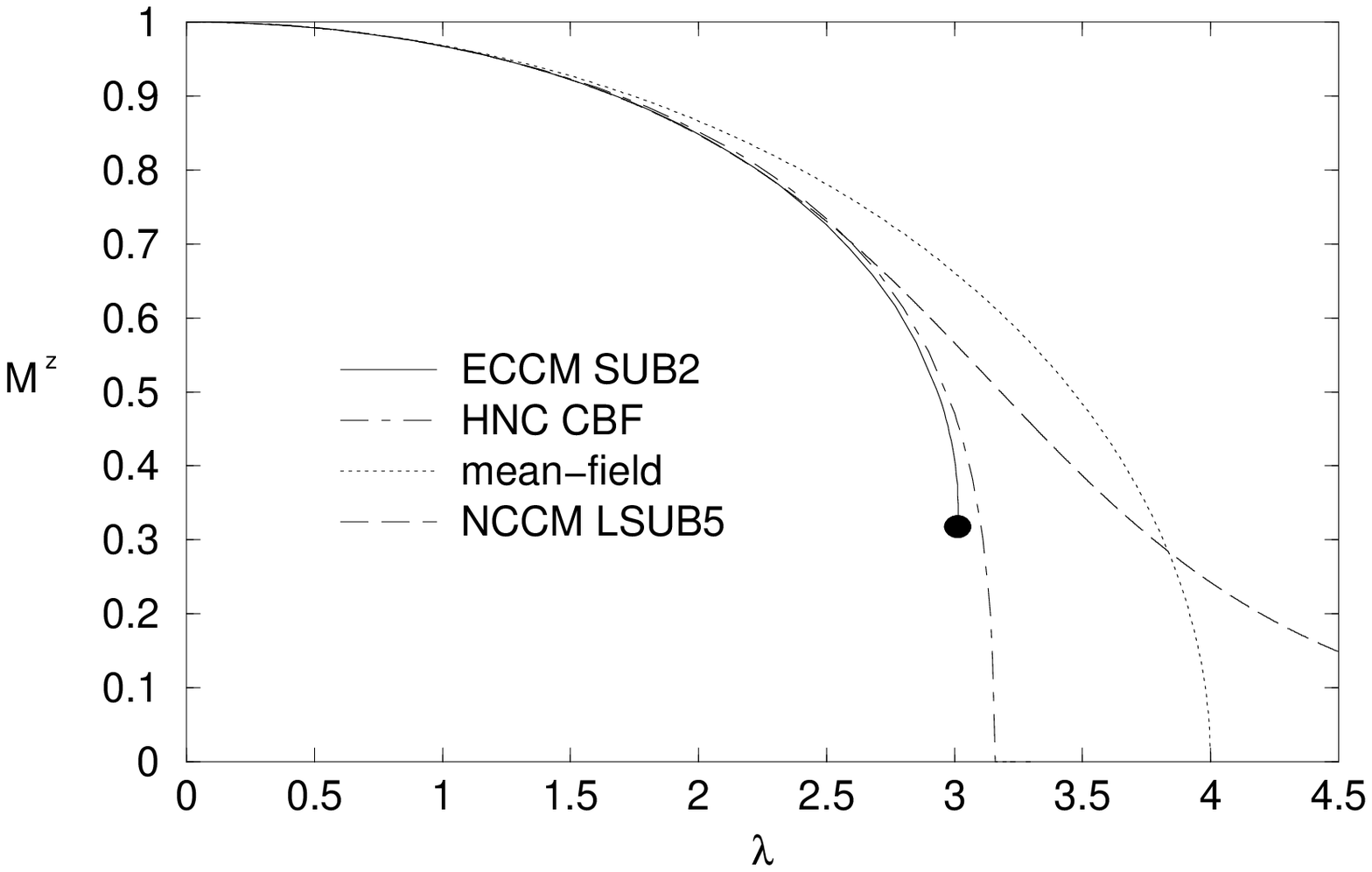}} 
   \caption{The longitudinal magnetization, $M^z$, as a function of 
the transverse magnetic field, $\lambda$, on the for the 
linear-chain case from the ECCM 
SUB2-$n$ scheme with the paramagnetic model state, 
the highest-order NCCM LSUB$n$ scheme with the ferromagnetic model 
state, a mean-field calculation, and the exact solution. The
LSUB$n$ truncation scheme retains only multiconfigurational 
creation operators in a localized area of $n$ contiguous sites.}
\label{M_l_2D}
\end{figure}

\begin{figure}[h]
   \epsfxsize=9cm 
   \centerline{\epsffile{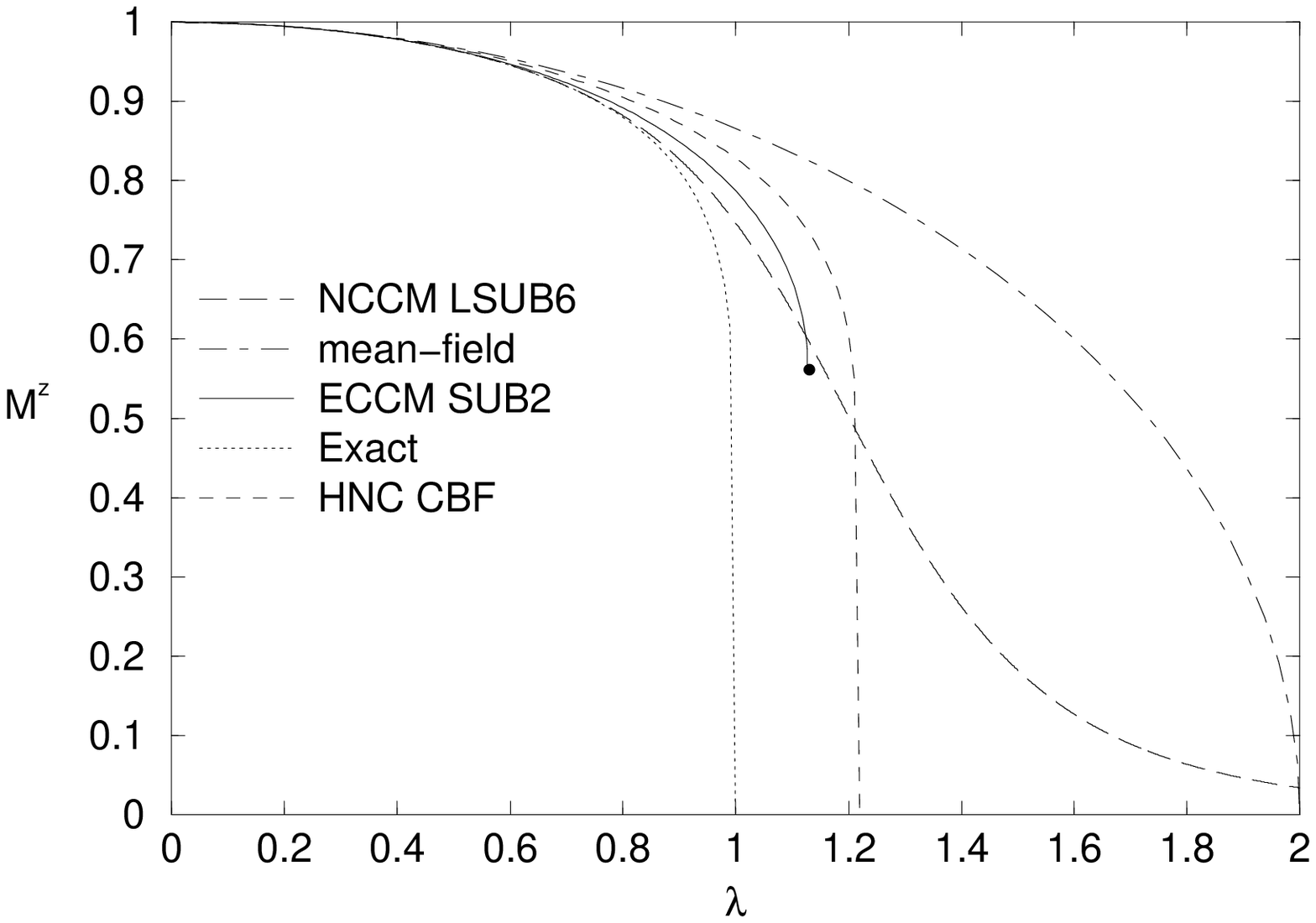}} 
   \caption{The longitudinal magnetization, $M^z$, as a function of 
the transverse magnetic field, $\lambda$, on the for the 
linear-chain case from the ECCM 
SUB2-$n$ scheme with the paramagnetic model state, 
the highest-order NCCM LSUB$n$ scheme with the ferromagnetic model 
state, a mean-field calculation, and the exact solution.}
\label{M_l_1D}
\end{figure}

The results from the square-lattice case (see Fig.~\ref{M_l_2D} 
and Table~\ref{2d_lc}) 
are in greater agreement with those from other methods
than the linear-chain results (see Fig.~\ref{M_l_1D} and Table~\ref{1d_lc}) 
are with the exact results. Although this is to be expected,
as the large quantum fluctuations on the linear chain  is numerically difficult to deal with.
 Our most accurate means of determining
the critical coupling constant is with the paramagnetic model state
starting from the ferromagnetic limit $\lambda = 0$. It yields a
value of $\lambda_c = 3.014$ which is close to the best high-temperature
series expansion results  \cite{HHO90} of $\lambda_c=3.0441(4)$ and the
best low-temperature series expansion results \cite{OHW91} of $\lambda_c=3.041(3)$.
Note that the series expansions, respectively, use the ferromagnetic 
and the paramagnetic model state. Other methods find similar, although less accurate 
results. \cite{Damian} The details are summarized in Table~\ref{2d_lc}.

\begin{figure}[h]
   \epsfxsize=9cm 
   \centerline{\epsffile{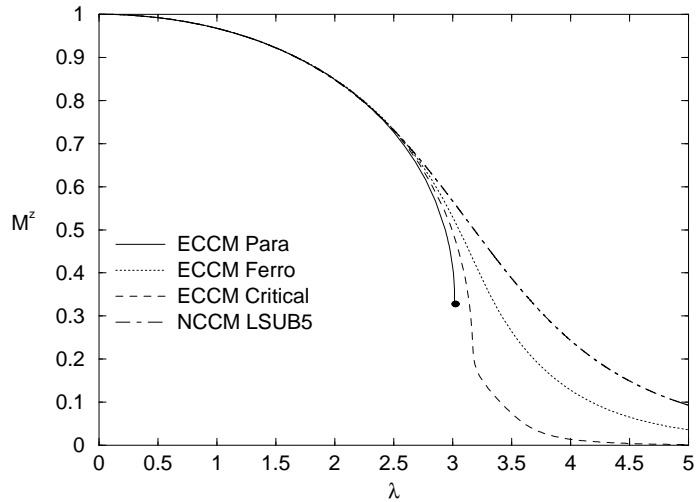}} 
   \caption{The longitudinal magnetization, $M^z$, 
as a function of the transverse magnetic field, $\lambda$, on the 
square-lattice case from the ECCM SUB2-$n$ scheme with the paramagnetic model state, a critical 
canted model state, which allows the solution to be 
continuous, and the ferromagnetic model state.
}
\label{magsols_2D}
\end{figure}

To investigate the model-state dependence further, we look at angles 
$\theta$ in between the ferromagnetic angle $\theta = 0$ and the 
paramagnetic angle $\theta = -\frac{\pi}{2}$. The largest angle at 
which the $M^z$ does not jump is rather close to the ferromagnetic angle.
The $M^z$ for these different model states are shown in Fig.~\ref{magsols_2D}. 
Note that this solution continues to $\lambda\to \infty$
because the $Z(2)$ symmetry of the model state is broken.
A continuous solution that traverses across the phase transition might be
in high demand, however, the solutions, from a canted angle model state,
do not have the right properties as the
excitation energy becomes complex before the phase-transition point.

\begin{figure}[h]
   \epsfxsize=9cm 
   \centerline{\epsffile{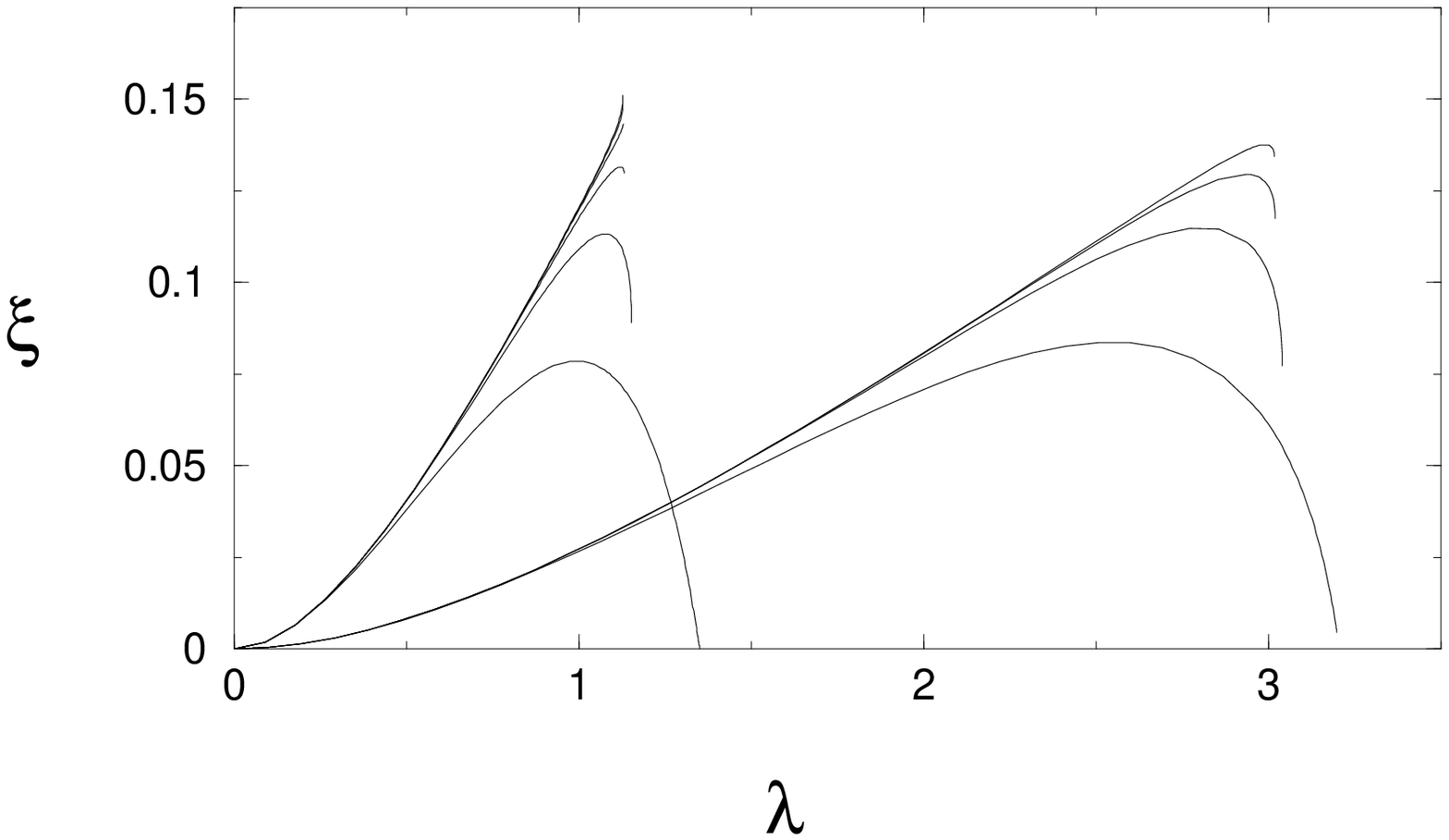}} 
   \caption{The correlation length $\xi$ as function of the coupling constant $\lambda$,
for both the linear-chain case (left) and the square-lattice case (right) with increasing
orders of approximation. For the linear-chain case the correlation length is converged, however,
not completely so for the square-lattice case. For the square-lattice case, 
each line
represents an approximation, where all correlators in a square, 
with lengths: 1, 2, 3, and 4, are taken into account, 
corresponding to 2, 5, 9, and 14 
distinct correlators.
}
\label{cor_1_2}
\end{figure}

The correlation length increases in the vicinity of the phase transition (see Fig. 
\ref{cor_1_2}), and shows good convergence. However, whether the 
correlation length is diverging at the phase transition is not clear. 
For the linear-chain case there are some indications of a sudden increase.
More interestingly, for lower-order approximations, the maximum 
correlation length occurs before the termination point, and, as the 
order increases, the termination point and the extremum in the correlation
length converge.

\section{Conclusions}

The purpose of studying the transverse Ising model with 
the ECCM was to gain further knowledge of the optimal 
approach for its implementation before tackling more 
complicated physical models. 
Two major considerations came to light: the 
pivotal importance of the model state in the
CCM calculation 
and that, importantly, the ECCM is the version of the CCM
which can show symmetry breaking. Encouragingly, the 
ECCM SUB2 numerical results 
are accurate in comparison to other numerical 
techniques and is in agreement with the series expansion 
result \cite{HHO90,OHW91} for $\lambda_c$ on the square lattice.

The calculations in this paper necessitates the inclusion of the
full ECCM SUB1 part of the functional, something we avoided in 
earlier work \cite{Josi2}
because of the computational difficulties. However, the advantages of
this inclusion are clear, since it allows us to study both phases in
one consistent approach. Unfortunately, the order of approximation to 
which this calculation can be pursued is lower, because of the greater
complexity. However, in the case of the transverse Ising model high-order 
approximations are not important since the convergence is
good, and apart from the results for correlation length we have 
fully converged results within the SUB2 approximation, at SUB2-20. 

The fact that the optimal ECCM SUB2 solution terminates 
in the critical region reveals that further improvements 
are necessary. Using more 
complicated model states, {\it i.e.}, a dimerized paramagnetic 
model state, could improve the situation. It seems that ECCM
can only be implemented efficiently in the SUB2 truncation scheme,
which is a serious drawback.
However, a new approximation scheme has recently been 
developed, \cite{thesis} which employs
``block-spin model states'' within the auspices of the SUB2 
scheme. Advantageously, this scheme can be 
viewed as a way of performing a re-summation of the 
diagrams arising from the usual single-spin model states 
and as such can allow the inclusion of a great-many 
correlations. The formulation of an algorithm \cite{thesis} 
for this scheme further increases its viability via 
computerization.

\section*{Acknowledgments}
One of us (J.R.) gratefully acknowledges support for this work in the form 
of a research grant from the Engineering and Physical Sciences Research 
Council (EPSRC) of Great Britain. We like to thank Damian Farnell 
for the use of his NCCM, CBF, and VQMC results,
and one of us (N.E.L.) likes to thank Niels Walet for some useful discussions.

\begin{table}[h]
\caption{A summary of the main features for the square-lattice case, 
with different model states and different starting points. The qualitative
difference with the linear-chain case is only for the excitation energy for 
the paramagnetic model state starting from large $\lambda$.}
\begin{center}
\begin{tabular}{|c|c|c|}
starting value & paramagnetic model state & ferromagnetic model state \\ \hline
$\lambda=0$ & solution terminates &  solution continues   \\
$\lambda \uparrow$ & $E_x$ vanishes &  $E_x$ turns complex   \\ 
 & $M^z$ terminates & $M^z$  decreases  \\ \hline
$\lambda=\infty$ & solution continues &  solution continues \\
$\lambda \downarrow$ & $E_x$ vanishes &  $E_x$ turns complex  \\ 
 & $M^z=0$ & $M^z$  increases  \\ 
 \end{tabular}
 \end{center}
 \label{re_tab}
 \end{table}

 \begin{table}[h]

 \caption{A comparison of the critical value of the 
 external magnetization at which various techniques 
 determine the presence of a phase transition on the 
 square lattice.}

\begin{center}
\begin{tabular}{cccccc}
   &  VQMC$^{\rm a}$ & HNC CBF$^{\rm a}$ & HT Series Expansion$^{\rm b}$ &
   LT Series Expansion$^{\rm c}$ & ECCM SUB2  \\ \hline
 $\lambda_c$& $3.15\pm 0.05$ & $3.12$ &$3.0441(4)$ & 3.041(3) & $3.014$  \\ \hline
 \end{tabular}
 \end{center}
${\ }^{\rm a}$ Reference \cite{Damian}\\
${\ }^{\rm b}$ Reference \cite{HHO90}\\
${\ }^{\rm c}$ Reference \cite{OHW91}\\
 \label{2d_lc}
 \end{table}

\begin{table}[h]
\caption{A comparison of the critical value of the 
external magnetization at which various techniques 
determine the presence of a phase transition on the 
linear chain.}
\begin{center}
\begin{tabular}{ccccc} 
  & VQMC$^{\rm a}$ & HNC CBF$^{\rm a}$ & Exact & ECCM SUB2  \cr \hline
   $\lambda_c$ & $1.206$ & $1.22$  & $1.0$ & $1.12$  \cr 
  \end{tabular}
 \end{center}
${\ }^{\rm a}$ Reference \cite{Damian}\\
 \label{1d_lc}
 \end{table}

\end{document}